# Magnetic Moment Peculiarities in Bismuth Bicrystals with Large Crystallite Disorientation Angles

F. M. Muntyanu[1,2], A. Gilewski[2*], K. Nenkov[2,4] J. Warchulska[2], and A. J. Zaleski[3]

[1]*Institute of Applied Physics, Academy of Sciences of Moldova, MD-2028, Chisinau, Moldova*

[2]*International Laboratory of High Magnetic Fields and Low Temperatures, PL-53-421, Wroclaw, Poland*

[3]*Institute of Low Temperatures and Structural Research, Polish Academy of Sciences, Wroclaw, Poland*

[4]*Leibniz-Institut fur Festkorper und Werkstofforschung, Dresden, Germany*

Magnetization measurements prove that the magnetic properties of large-angle ($\theta > 30°$) bismuth bicrystals with crystallite interface (CI) of twisting type essentially differ from well-known results on single-crystalline specimens. Two superconducting phases were observed at CI of bicrystals (ordinary rhombohedral Bi is not a superconductor). It was shown that one of them (phase with $T_c \sim 8,4$ K) is localized in central part of crystallite interface. It was also found that in adjacent layers (width of layer $d_2 \sim 20$ nm) of CI the proximity effect is significant.

*Introduction.* - Magnetic properties of bismuth are intensively studied for a long time. This interest is caused by two reasons. First, in bismuth the de Haas-van Alphen effect was found [1], which later was used for the Fermi surface determination almost in all metals and alloys. Second, the monotonic part of magnetic susceptibility in Bi is by an order of magnitude higher than in metals. In spite of the fact that the doubled ratio of the spin to orbit splitting of the energy levels of electrons is close to 2 and according to the Pauli-Landau formula susceptibility $\chi = M / H$ must be above zero, Bi shows anomalous diamagnetism.

Quantitative calculation of Bi susceptibility [2, 3, 4, 5] has shown that the nearest bands make significant contribution to it, therefore $\chi$ considerably depends on the electron spectrum behavior on large momentum. The field dependences of the magnetic moment $M$ were studied basing on different models of the electron energy spectrum. A satisfactory agreement of the theoretical calculations with the experimental data is obtained [6, 7]. It was found in [5] that the magnetic susceptibility component $\chi_\parallel (H \parallel C_3)$ in Bi is determined by electrons and holes in filled band and practically does not depend on temperature. According to [5] at $H \perp C_3$ the susceptibility consists of two components $\chi^1_\perp$ (depends on temperature, magnetic field and parameters characterizing the charge carriers in $L$-point of the Brillouin zone) and $\chi^2_\perp$ (weakly depends on temperature and field value). Charge carriers in Bi energy bands in weak magnetic fields give paramagnetic contribution, hence when their concentration is high the diamagnetism decreases with the chemical potential increasing. The susceptibility as a function of the chemical potential has peculiarities connected to the Fermi surface topology, which become smooth at finite temperatures, when the interaction of the charge carriers with each other or with impurities is intensified [8]. These processes must manifest itself most clearly in bicrystal samples, where, in the region of inner boundary, significant changes in the electron energy spectrum take place [9, 10, 11], the charge carrier density is increased, the charge carrier interaction with the inner boundary vary, etc.

In the present work it is shown that magnetic properties of Bi bicrystals differ considerably from the known data [6, 7] for single crystal samples (superconducting properties are shown, charge carrier paramagnetism is intensified, temperature and field dependences of the magnetic susceptibility changes, etc.).

Bi bicrystals are obtained by the zone recrystallization method using double seed technique. Samples for measurements were prepared in the form of parallelepipeds, where CI with the width of about 100 nm was normal to the preferential plane of perfect cleavage.

We have studied two types of bicrystals, presenting itself: a) two crystallites of rhombohedral structure with the layers laid perpendicular to the axis $C_3$ [12], b) two crystallites, in one of which the layers are laid perpendicular to the axis $C_3$, and in the other the normal to the layer plane and direction of the first crystallite axis $C_3$ make up an angle of about 30°. The second type of the bicrystals had *p*-type of conductivity and was of a special interest because the CI exhibits superconducting properties (for some samples $T_{onset}$ ~ 16 K [10]). The magnetic moment of these bicrystals was studied in the temperature range (1,8 - 300) K and in the fields up 70 kOe using Cahn balance, SQUID magnetometer (Quantum Design) and Physical Property Measurement System (PPMS). The measurements were carried out in the International Laboratory of High Magnetic Fields and Low Temperatures (Wroclaw, Poland), Institute of Low Temperatures and Structure Research of the Polish Academy of Sciences in Wroclaw (Poland), Institute for Solid State and Material Research (Dresden, Germany).

*Temperature dependence of magnetization.* - Figure1 shows examples of temperature dependences of the static magnetic moment of Bi bicrystals with the CI of the twisting type and with large angle of disorientation of the crystallites. At low temperatures two superconducting phases are observed: one has $T_c$ ~ 8,4 K, and the other $T_c$ ~ 4,3 K.

The existence of superconducting phases with different critical temperatures means that at CI, alongside with other phenomena (changes of the electron-phonon interaction, electron-electron repulsion, free path length of charge carriers, etc.), also, appreciable changes of density of Cooper pairs at distances comparable with the coherence length are taking place.

Superconducting phase with the critical temperature $T_c \sim 8,4$ K is very likely localized in the central part of CI and phase with $T_c \sim 4,3$ K- in adjacent layers. Phase with $T_c \sim 8,4$ K was earlier described in detail in [9, 10], where it is shown, that the coherence length $\xi(0) \sim 12$ nm being much less than width of the central part of CI ($d_1 \sim 60$ nm), so the adjacent layers do not improve conditions for correlation of the Cooper pairs in the central part of the crystallite interface.

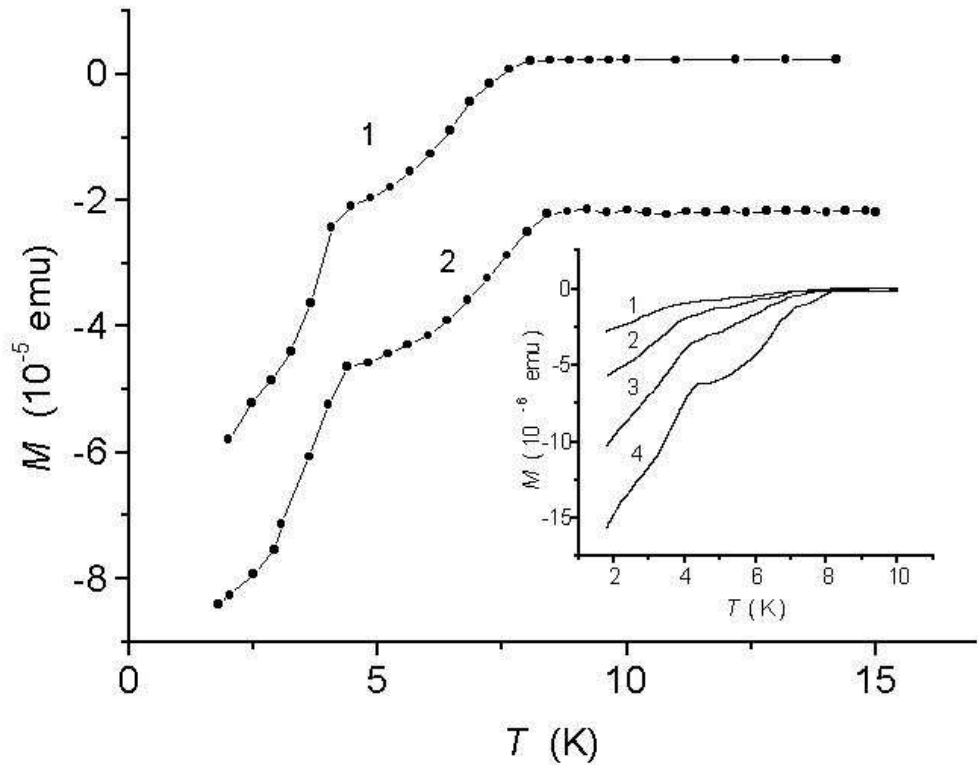

FIG. 1. Temperature dependences of magnetic moment of Bi bicrystals with the CI of the twisting type. 1 - $\theta_1 = 290$, $\theta_2 = 110$; 2 - $\theta_1 = 620$, $\theta_2 = 20$; 1 - 10 Oe; 2 - 20 Oe. Inset: Temperature dependences of magnetic moment at different applied magnetic fields in Bi bicrystal with $\theta_1 = 290$, $\theta_2 = 110$; 1 - 1000 Oe, 2 - 300 Oe, 3 - 100 Oe, 4 - 0 Oe.

On the other hand, thickness of the adjacent layers of bicrystals CI was estimated to be $d_2 \sim 20$ nm (width of layers as determined by the field when magnetoresistance quantum oscillations in normal state start manifesting themselves). They border Bi normal phase (width of crystallites greatly exceed the coherence length). The charge carrier concentration in these layers is approximately twice less than in the central layer and is by three orders of magnitude higher than in crystallites. In this connection, role of the proximity effects and increase of the free path length of the charge carriers (with the free path length the coherence length increases and vice versa) may be significant.

The magnitude of the relative change of the superconductor transition temperature in the adjacent layers on the boundary with the normal phase may be determined [13, 14] from the following relation:

$$[T_{cs} - T_c(d_2)] / T_{cs} = (\pi^2 / 16) (\hbar^2 / m\alpha) d_2^{-2}, \tag{1}$$

where: $m\alpha \sim \pi^2/4 \, (t/d_2)^2$, $t \sim 0{,}6\xi(0)$ (for pure superconductor), $T_{cs}$ is the critical temperature of the superconducting transition of the adjacent layer on the boundary with vacuum.

Estimations show that if the coherence length $\xi(0) \sim 12$ nm is used [9], then $T_{cs}$ of the adjacent layer due to influence of bismuth normal phase (crystallites) is decreased by about 2 K (that is $T_{cs} \sim 6{,}3$ K). However, if we suppose that in the adjacent layer the given value of $\xi(0)$ is about 15% higher (the supposition is quite grounded taking into account that the charge carrier concentration in the adjacent layers is almost twice less than in the central layer), then the critical temperature $T_{cs}$ grows up to values characteristic of the central layer.

From the temperature dependences of the AC magnetic moment measured in the magnetic field (Fig.1, Inset) the upper critical field $H_{c2}(T)$ is determined for the superconducting phase with $T_c \sim 4{,}3$ K (for the phase with $T_c \sim 8{,}4$ K these parameters are determined in [9]). By the slope of dependence $H_{c2}(T)$ the temperature derivative of the upper critical field is estimated and it is found that $dH_{c2}/dT \sim 5{,}5$ kOe. The estimations of $H_{c2}(0)$ were carried out by the formula [15]:

$$H_{c2}(0) = -0{,}69\, T_c\, (dH_{c2}/dT). \tag{2}$$

As a result, it is found that $H_{c2}(0) \sim 16{,}6$ kOe, and the coherence length $\xi(0) \sim 14$ nm (it is determined from the relation $\xi^2(0) = \Phi_0 / 2\pi H_{c2}(0)$, where $\Phi_0 = 2{,}07 \times 10^{-7}$ G cm$^2$).

*Field dependence of magnetization.* - According to [4] in single crystal bismuth for $H \parallel C_3$ the magnetic moment changes almost linearly in the whole fields range (except for the lowest ones, the small changes are ascribed connected to the paramagnetic contribution of the impurity charge carriers), being all time diamagnetic.

Figure 2 shows the magnetic moment field dependences in large-angle Bi bicrystals at the magnetic field orientation along the inner boundary plane (in crystallites this direction corresponds to $H \parallel C_3$). As it is seen from the figure, the $M(H)$ curves are characterized by pronounced diamagnetism in weak fields (for temperatures $T < T_c$ the magnetic susceptibility $|\chi|$ sharply increases (Fig.2, Inset *a*) when the magnetic field decreases), paramagnetic maximum (in the magnetic field region 0,5 - 2 kOe), weak dependence on temperature and almost linear change at $H > 2$ kOe (Fig.2, Inset *b*). The diamagnetism increase in Bi bicrystals at $H < 0{,}5$ kOe and low temperatures is explained not only by disappearance of the paramagnetic contribution of the impurity carriers, but also by the CI transition into the superconducting state. Observation of the well pronounced Meissner effect confirms the fact that on the inner boundary there is the superconducting material in the amount sufficient for influence on the magnetic moment value.

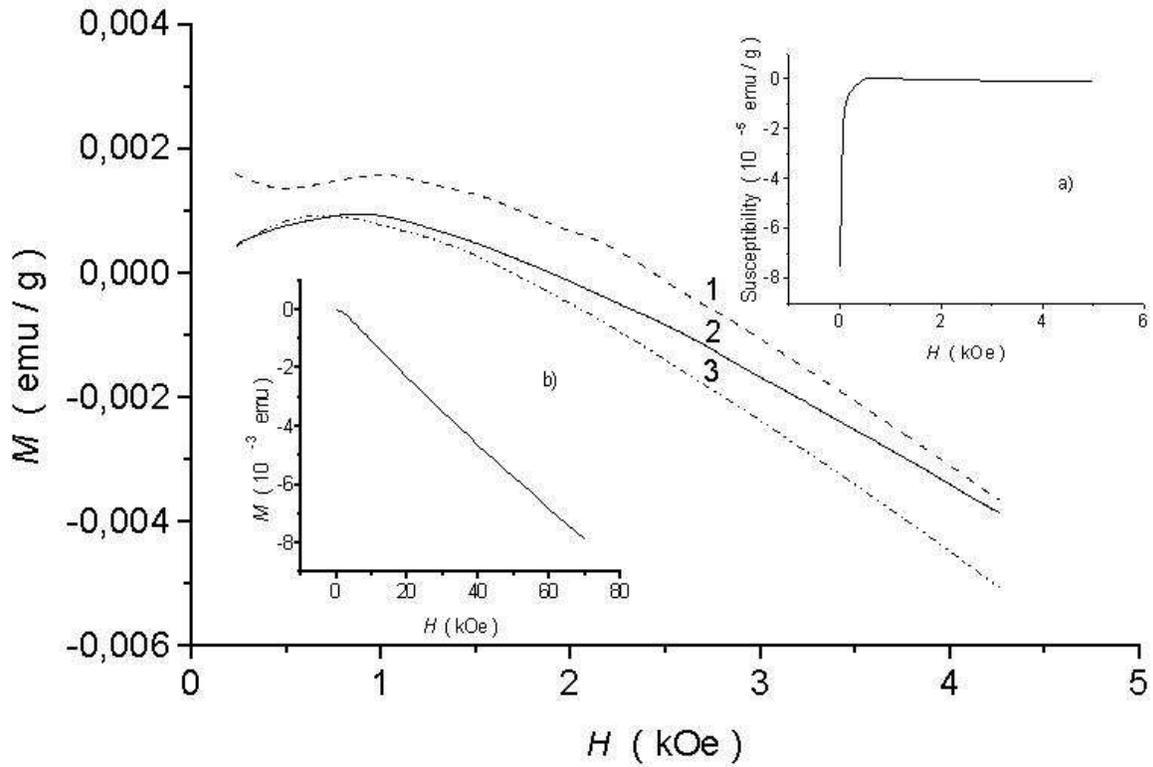

FIG. 2. Field dependences of magnetic moment of Bi bicrystal with CI of the twisting type ($\theta_1 = 62^0$, $\theta_2 = 2^0$). 1 - 4,2 K; 2 - 300 K; 3 - 77 K. Inset (a): Field dependence of magnetic susceptibility at $T = 1,8$ K of Bi bicrystals with $\theta_1 = 62^0$, $\theta_2 = 2^0$. Inset (b): Field dependence of magnetic moment at $T = 2$ K of Bi bicrystal with $\theta_1 = 29^0$, $\theta_2 = 11^0$.

On the other hand, concentration of the main charge carriers in CI of Bi bicrystal is by about three orders of magnitude higher than in crystallites, so their paramagnetic contribution (in the normal state of the samples) must be significant and rapidly disappear with the field increase (disappearance of the paramagnetic contribution of these carriers takes place in large fields due to their higher concentration). In the magnetic field the charge carrier contribution to the magnetic moment of crystallites qualitatively changes too (from the paramagnetic in weak fields to the diamagnetic in strong fields) [5]. Therefore, in strong fields diamagnetism in large-angle Bi bicrystals will increase ($M_H \sim H$), the response of the system with magnetism conditioned by the conduction electrons must not depend appreciably on temperature (the magnetic moment of the conduction electrons is proportional to the density of states on the Fermi level) and on parameters characterizing the charge carriers.

Figure 3 shows the results of measurement of the magnetization loop at 2 K in one of the samples of large-angle bicrystals with CI of twisting type.

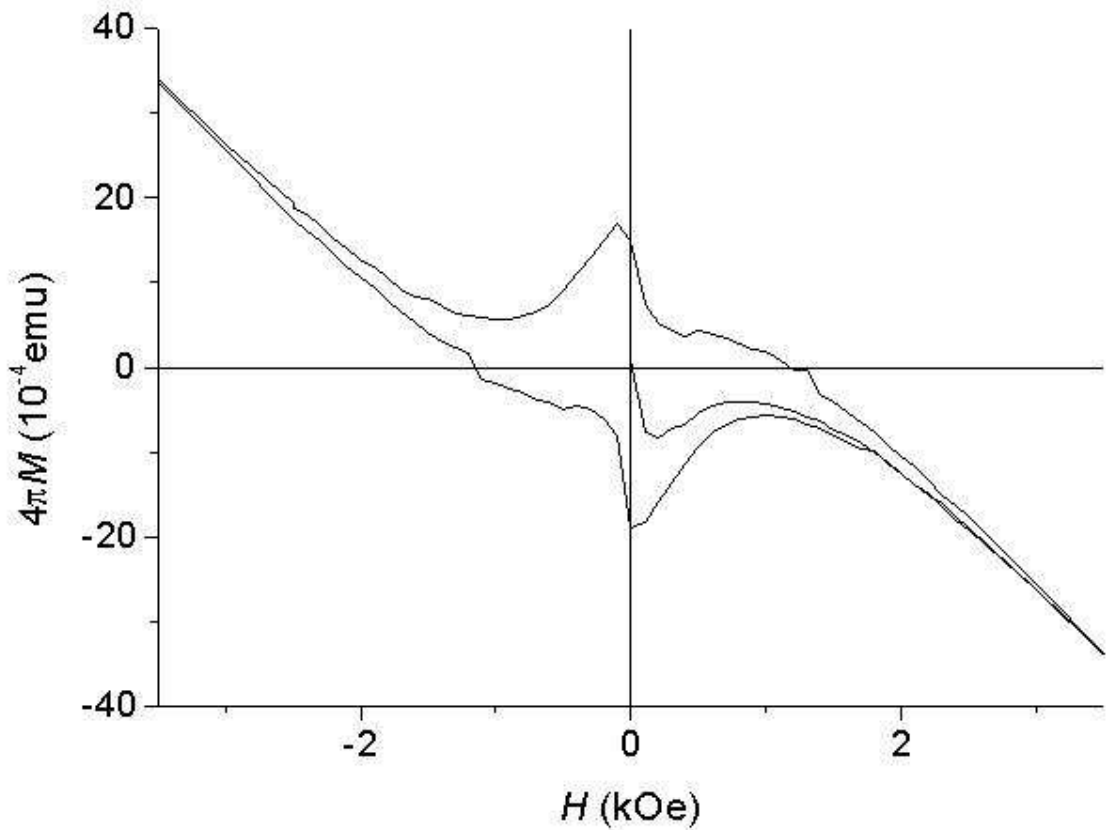

FIG. 3.   Magnetic hysteresis loop at $T = 2$ K of Bi bicrystal with $\theta_1 = 29^0$, $\theta_2 = 11^0$.

The magnetization loop clearly shows the behavior typical for strong type-II superconductor. From the width of the hysteresis loop the critical current density in superconducting regions of the sample was estimated, using the result of [16] that

$J_c(H) = (10/4\pi)(m^+ - m^-)/w$, where $w$ is the thickness of the CI and $m^+ = 4\pi M^+(H)$, $m^- = 4\pi M^-(H)$, $M^+(H)$, $M^-(H)$ denote the positive and negative parts of magnetization. It is obtained that the critical current density at 1,13 kOe is $J_c \sim 1{,}2 \times 10^3$ A cm$^{-2}$. According to the data shown in this figure, the magnetic field begins to penetrate partially into the superconductor thickness at $H_{c1} \sim 130$ Oe. The measured values of the critical magnetic fields $H_{c1}$ and $H_{c2}$ testify to the fact that the penetration depth $\lambda(0) \sim 150$ nm.

Thus, the obtained data confirm existence of two superconducting phases on the twisting inner boundary in large-angle bismuth bicrystals, which might be associated with the central part (the phase with $T_c \sim 8{,}4$ K) and the adjacent layers (the phase with $T_c \sim 4{,}3$ K) of the crystallite interface. These phases are close in structure, but they differ in electron properties in both superconducting and normal states.

*Conclusions.* - We have investigated magnetic properties of large-angle bismuth bicrystals with crystallite interface of twisting type and have found two new superconducting phases with different critical temperatures ($T_c \sim 8{,}4$ K and $T_c \sim 4{,}3$ K), the magnetic susceptibility $|\chi|$ increases in weak magnetic fields and appearance of the paramagnetic maximum on dependences $M(H)$ in the field range 0,5 - 2 kOe.

Observation of two new (stable under atmospheric pressure) superconducting phases on the inner boundary in large-angle Bi bicrystals is a result of structural reconstruction of the rhombohedral crystal structure *A7* under the influence of twisting deformations. This modification of bismuth differs in parameters from high pressure phases Bi11-Bi1V [16], metastable phases [17] obtained in ultrathin layers, and granulated systems consisting of rhombohedral clusters [18].

[a]

---

[a] *Corresponding author. Electronic address: gil@alpha.ml.pan.wroc.pl

[1]  W.Y. de Haas and P.M. van Alphen, Proc. Acad. Sci. (Amsterdam) **33**, 1106 (1930).

[2]  E.N. Adams, Phys. Rev. **89**, 633 (1960).

[3]  F.A. Buot and J.M. McClure, Phys. Rev. **6**, 4525 (1972).

[4]  J.M. McClure and D. Shoenberg, J. of Low Temp. Phys. **22** nr ¾, 233 (1976).

[5]  S.D. Beneslavsky and L.A. Falkovsky, Zh. Eksp. Teor. Fiz. **69**, 1063 (1975).

[6]  L. Wehrli, Phys. Condens. Mater. **8**, 87 (1968).

[7]  B.I. Verkin, L.B. Kuzmitheva, and I.V. Svethkarev, Zh. Eksp. Teor. Fiz., Pisma, **6,** 757 (1967).

[8]  N.B. Brandt, M.V. Semenov and L.A. Falkovsky, J. of Low Temp. Phys. **27** nr 1/2, 75 (1977).

[9]  D.V. Gitsu, A.D. Grozav, V.G. Kistol, N.I. Leporda and F.M. Muntyanu, Zh. Eksp. Teor. Fiz., Pisma, **55**, 389 (1992).

[10]  F.M. Muntyanu and N.I. Leporda, Solid State Phys. **37**, 549 (1995).

[11]  S.N. Burmistrov and L.B. Dubovskii, Phys. Lett. A **127**,120 (1988).

[12]  F.M. Muntyanu, Iu.A. Dubkovetskii and A. Gilewski, Solid State Phys. **46,** 1763 (2004).

[13]  N.R. Werthamer, Phys. Rev. **132**, 6, 2440 (1963).

[14]  A.A. Abrikosov, *Osnovy teorii metallov* (Nauka, Moskva, 1987), p 520.

[15]  N.R. Werthamer, E. Helfand and P.C. Hohenberg, Phys. Rev. **147**, 295 (1966).

[16]  E.M. Savitskii, O. Henkel, and Yu.V. Efremov, *Physics and Chemistry of the Synthesis of Superconducting Materials* (Metallurgiya, Moscow, 1981).

[17]  B.G. Lazarev, E.E. Semenenko, V.I. Tutov and A.A. Chupikov, Fizika Nizkich Temperatur **4**, 957 (1978).

[18]  B. Weitzel and H. Micklitz, Phys. Rev. Lett. **66**, 385 (1991).